\documentclass[aps,prb,showpacs,floatfix,twocolumn]{revtex4}
\usepackage{amssymb} 
\usepackage{amsmath}
\usepackage{graphicx}
\usepackage[latin1]{inputenc}
\begin{document}
\setlength{\parindent}{0pt} 
\newcommand{\llangle}{\langle \langle}
\newcommand{\rrangle}{\rangle \rangle}
\title{Improved position measurement of nanoelectromechanical systems using
  cross correlations} 
\author{C.B. Doiron} 
\affiliation{Department of Physics and Astronomy, University of Basel, 
CH-4056 Basel, Switzerland}
\author{B. Trauzettel} 
\affiliation{Department of Physics and Astronomy, University of Basel, 
CH-4056 Basel, Switzerland}
\author{C. Bruder} 
\affiliation{Department of Physics and Astronomy, University of Basel, 
CH-4056 Basel, Switzerland}


\begin{abstract}
We consider position measurements using the cross-correlated output of two
tunnel junction position detectors. Using a fully quantum treatment, we
calculate the equation of motion for the density matrix of the coupled
detector-detector-mechanical oscillator system. After discussing the presence
of a bound on the peak-to-background ratio in a position measurement using a
single detector, we show how one can use detector cross correlations to
overcome this bound. We analyze two different possible experimental
realizations of the cross correlation measurement and show that in both cases
the maximum cross-correlated output is obtained when using twin detectors and
applying equal bias to each tunnel junction. Furthermore, we show how the
double-detector setup can be exploited to drastically reduce the added
displacement noise of the oscillator.
\end{abstract}

\date{August 2007} \maketitle

\section{Introduction}
It is expected that, in the near future, position measurements of
nanomechanical systems will reach the quantum limit. Experimental
progress in this direction is very fast and displacement sensitivities
near the standard quantum limit have already been
demonstrated\cite{knobel2003,lahaye2004,naik2006}.

In the current generation of experiments, the coupling between the
resonator and the mesoscopic detector is typically very weak. The
position measurement can therefore \emph{not} be seen as a strong
projective measurement. It is better
described within the framework of weak measurement theory that was
recently developed in the context of solid-state quantum
computing\cite{averin2003,korotkov2001,pilgram2002,clerk2003}. This theory
describes a continuous measurement process where the information about
the measured object can be extracted, for instance, from the spectral
density of the detector (and not simply from its average output). An
important result in this theory is the Korotkov-Averin bound, which
puts an upper limit of 4 to the ratio of the contribution of the
measured state to the detector's spectral density, and the intrinsic
background detector noise, for any linear detector measuring a
two-level system.

Since a quantum position measurement by a mesoscopic detector can
be described within the same theoretical framework as a qubit
measurement, one might ask if such a bound also exists in the case of
a position measurement. In this article, we first show that, for fixed system
parameters, the 
peak-to-background ratio in the spectral density of a position
detector weakly coupled to an oscillator is also bounded from
above. This result is obtained by considering the example of a single
tunnel-junction detector, a simple detector that has been thoroughly
studied theoretically\cite{bocko1988,
yurke1990,schwabe1995,mozyrsky2002,clerk2004b,wabnig2005,wabnig2007}
and realized
experimentally\cite{cleland2002,flowers-jacobs2007}. 

\begin{figure}[htbp]
\begin{center}
\includegraphics[scale=0.45]{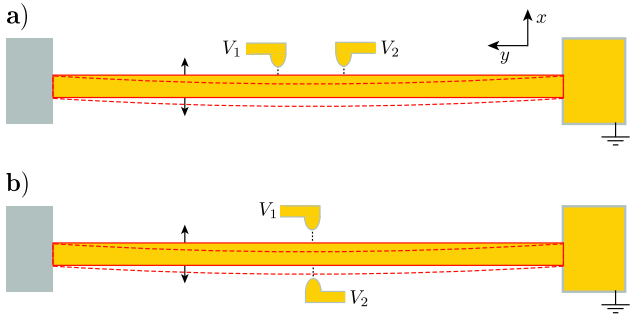}
\caption{ 
(Color online) The two typical detector configurations examined in this
article. In both cases, the movement of the oscillator is along the $x$ direction in the $xy$ plane, as depicted by the $\leftrightarrow$ sign.
a) \emph{In-phase configuration}, where two detectors (with bias $V_1$ and
$V_2$, respectively) are
located on the same side of the central part of the oscillator, such
that both detectors couple in the same way to the position of the
oscillator. This is covered in Sec. \ref{case 1}. b)
\emph{Out-of-phase configuration}, where the detectors are located on
each side of the oscillator.
When the position of the oscillator is such that the tunneling amplitude of
one junction is increased, the tunneling amplitude of the other junction is
therefore decreased. This is covered in Sec. \ref{case 2}.}
\label{Setup}
\end{center}
\end{figure}

Besides showing that the peak-to-background ratio is bounded in the
typical single-detector position measurement, we also propose, in this
article, two simple experimental configurations (Fig.~\ref{Setup})
where, by using the cross correlations between two detectors, the bound on the
peak-to-background ratio can be overcome. As the 
oscillator-independent parts of the output signal of the two detectors
are uncorrelated, the background noise in these configurations is zero
and therefore the peak-to-background ratio diverges.  In the context of qubit
readout, this idea has already been proposed in an insightful work by Jordan
and B{\"u}ttiker \cite{jordan2005} and was shown experimentally to improve
readout fidelity\cite{buehler2003}. Experimentally, position measurements
should hence also profit from using cross-correlated detector outputs. We
analyze in detail the two configurations presented in Fig.~\ref{Setup} and
obtain analytical results for the optimal cross-correlated signal as a
function of different detector parameters.

Previous studies\cite{clerk2004a} of the position measurement problem focused
on finding the conditions for quantum-limited detector sensitivity,
under which one minimizes the total detector contribution to the output displacement noise. We show that the
double-detector setup proposed here can in fact be used to almost totally get
rid off the added displacement
noise of the oscillator due to detector back-action. This is a remarkable result that nicely complements
the general single-detector analysis made in Ref.~\onlinecite{clerk2004a}.

The article is organized as follows: in Section \ref{sec2}, we
introduce the formalism used in the rest of the paper, viz., a master
equation for the $m$-resolved density matrix, where $m$ is the number
of charges that have passed through the detector. This equation of
motion allows us to find expressions for the combined moments of
charge (detector) and oscillator quantities.
In Section \ref{sec3}, the formalism is applied to the case of one
position detector coupled to the oscillator.\cite{clerk2004b} 
We analyze the peak-to-background ratio and show that this quantity is
always bounded from above in the single-detector case. This bound cannot be made arbitrarily large simply by increasing the detector sensitivity.
Section \ref{sec4} generalizes this treatment to a configuration with
two detectors and demonstrates that measuring the current 
cross correlations of the two detectors allows one to get arbitrarily high
values of the peak-to-background ratio: i.e., it is possible to eliminate the bound that exist in the single-detector case. In Section \ref{sec5}, we demonstrate how the
proposed setup can be used to diminish the added position noise of the oscillator
induced by the presence of the detector, allowing position measurement 
beyond the standard quantum limit derived for a single detector.

\section{Equation of motion for the density matrix}
\label{sec2}
Approaches based on quantum master equations have proven useful in the
study of nanomechanical systems \cite{utami2004,rodrigues2005a}. By
writing the equation of motion for the density matrix of the full
(detector and oscillator) system and tracing out the detector degrees
of freedom, one can obtain an equation of motion for the reduced
density matrix describing the evolution of the oscillator taking into
account the coupling to the detector. In order to investigate
electronic transport in the coupled system, it is useful to refine
this approach to keep track of $m$, the number of charges that passed
through the detector. This allows one to calculate an equation of motion
for the $m$-resolved
density-matrix\cite{clerk2004b,rammer2004,wabnig2005}, a quantum
equivalent to the $m$-resolved master equation approach widely used in the study of transport properties of classical
nanomechanical systems\cite{armour2004,doiron2006}.

To study the current cross correlations between two tunnel junction
position detectors coupled to an oscillator, we use such a fully
quantum approach. We label the detectors with the index $\alpha =1,2
$ and model each of them as a pair of metallic leads with constant
density of states $\Lambda_\alpha$ (in the energy range relevant to
tunneling) coupled via the tunneling Hamiltonian $H_{\mathrm{tun}}$.
The Hamiltonian for one detector can therefore be written as a sum of
a bath Hamiltonian $H_{B,\alpha}$ describing the leads of junction
$\alpha$ and a tunneling Hamiltonian $H_{\mathrm{tun},\alpha}$
\begin{align}
H_{\mathrm{det},\alpha} &= H_{B,\alpha} + H_{\mathrm{tun},\alpha} \\
H_{B,\alpha}&= \sum_k \varepsilon_{k,\alpha} c^{\dagger} _{k,\alpha}
c_{k,\alpha} + \sum_q \varepsilon_{q,\alpha} c^\dagger _{q,\alpha}
c_{q,\alpha} \\ H_{\mathrm{tun},\alpha}&= T_\alpha (\hat{x})
Y^\dagger_\alpha \sum _{k,q} c^\dagger _{k,\alpha} c_{q,\alpha} +
T^\dagger_\alpha (\hat{x}) Y_\alpha\sum _{k,q} c^\dagger_{q,\alpha}
c_{k,\alpha}\;,
\end{align}
where $k(q)$ is a wave-vector in the right(left) lead. The coupling
between the detector and the position of the oscillator is modeled by
a linear $x-$dependence of the tunneling amplitude
\begin{align}
T_\alpha (\hat{x}) = \frac{1}{2\pi \Lambda_\alpha} \left (\tau_{0,\alpha}
+ e^{i \eta_\alpha} \tau_{1,\alpha} \hat{x} \right ) \;.
\end{align}
In this equation, $\tau_{0,\alpha}$ is the bare (oscillator-independent) tunneling amplitude of detector $\alpha$, $\hat{x}$ is
the position operator of the oscillator and $\tau_{1,\alpha}$ is the
part of the full tunneling amplitude 
detector $\alpha$ that depends on the position of the oscillator. We allow for a general relative phase
$\eta_\alpha$, describing the details of the coupling between the
tunnel junction and the oscillator. Such a phase can in principle be
controlled by a magnetic flux penetrating an extended tunnel junction
consisting of a loop containing two junctions, one of which couples to
the oscillator.  Note that in our notation $\tau_{0,\alpha}$ is
dimensionless and $\tau_{1,\alpha}$ has dimensions of one over length
and that we assume for simplicity that the tunneling amplitudes do not
depend on the single particle energies
$\varepsilon_{k,\alpha(q,\alpha)}$. The operator
$Y^{(\dagger)}_\alpha$ decreases (increases) $m_\alpha$, the number of
charges that tunneled through junction $\alpha$. Its presence in the
tunneling Hamiltonian allows one to keep track of the transport
processes that occur during the evolution of the system.

We are interested in calculating the equation of motion for the
reduced, $m_\alpha$-resolved, density matrix
\begin{align}
\rho (m_1,m_2;t) = \langle m_1, m_2 \lvert \rho _{\mathrm{osc}} \rvert
m_1, m_2 \rangle \;,
\end{align}
where $\rho_{\mathrm{osc}} = \mathrm{Tr}_{B} \{ \rho _{tot} \} $ is
the reduced density matrix that is obtained by tracing out the leads'
degrees of freedom from the full system density matrix. Within a
Born-Markov approximation, the equation of motion of
$\rho_{\mathrm{osc}}$ can be expressed as
\begin{align}\label{eq motion 2 detector setup} 
\begin{split}
\frac{\partial}{\partial t}& \rho_{\mathrm{osc}} (t) = -
\frac{i}{\hbar} [H_{\mathrm{osc}}, \rho_{\mathrm{osc}} (t)] \\&-
\frac{1}{\hbar^2} \int _{-\infty} ^0 d\bar{t} \, \mathrm{Tr}_{B} \{
\left[ H_{\mathrm{tun}} , \left[ H_{\mathrm{tun} } (\bar{t}\,) ,
\rho_{\mathrm{osc}} (t) \otimes \rho_{B}\right] \right] \} \;,
\end{split}
\end{align}
where $H_{\mathrm{tun}} = H_{\mathrm{tun},1}+H_{\mathrm{tun},2}$ is
the total tunneling Hamiltonian, the trace is on both pairs of leads,
$\rho_{B}$ is the coupled density matrix of the two sets of leads and
\begin{align}
H_{\mathrm{osc}} &= \hbar \Omega (\hat{a}^\dagger \hat{a} +1/2 ) =
\frac{\hat{p}^2}{2 M} + \frac{M \Omega^2 \hat{x}^2}{2} \;,\\
H_{\mathrm{tun} } (t) &= \sum _\alpha e^{i H_{0,\alpha} t / \hbar}
H_{\mathrm{tun},\alpha }e^{-i H_{0,\alpha} t / \hbar} \;.
\end{align}
with $H_{0,\alpha}=H_{\mathrm{osc}} +  H_{B,\alpha}$.
In our system, the Born approximation corresponds to assuming that
tunneling in both tunnel junctions is weak enough so that it can be
treated using second-order perturbation theory. 
The Markov approximation, on
the other hand, is valid as long as the typical correlation times in
the leads ($h/eV$) are much shorter than $2\pi/\Omega$, i.e. the typical
evolution time of the oscillator. In
practice, this limits the applicability of the following results to
the strongly biased case $eV \gg \hbar \Omega$. This is experimentally
feasible since typical oscillator frequencies $\Omega$ 
are between 10 -- 100 MHz and the measurements are done at a much larger bias
voltage than these frequencies. \cite{knobel2003,lahaye2004,naik2006,flowers-jacobs2007}

Since the leads of detector 1 are totally independent of those of
detector 2, $\rho_{B}$ can be written as a tensor product of the
density matrices describing each pair of leads $\rho_{B}=\rho_{B_1}
\otimes \rho_{B_2}$. Also, as $H_{\mathrm{tun},\alpha}$ has no
diagonal contribution in the basis that diagonalizes $H_{B,\alpha}$,
the trace over leads $\alpha$ of a quantity that is linear in
$H_{\mathrm{tun},\alpha}$ vanishes. As a result of those two
properties, the trace in Eq.~(\ref{eq motion 2 detector setup}) can be
rewritten as a sum over two traces, each involving only one pair of
leads
\begin{align}
\begin{split}
&\mathrm{Tr}_{B} \{ \left[ H_{\mathrm{tun}} , \left[ H_{\mathrm{tun} }
(\bar{t}\,) , \rho_{\mathrm{osc}} (t) \otimes \rho_{B}\right] \right]
\} \\ &= \sum_\alpha \mathrm{Tr}_{B_\alpha} \{ \left[
H_{\mathrm{tun,\alpha}} , \left[ H_{\mathrm{tun,\alpha} } (\bar{t}\,)
, \rho_{\mathrm{osc}} (t) \otimes \rho_{B_\alpha}\right] \right] \}
\;.\end{split} 
\end{align} 
This effectively makes the two-detector
problem two single-detector problems. The trace over the leads' 
degrees of freedom is then carried out in the standard
way. \cite{blum1996} 

As mentioned above, we are interested in calculating the
time-evolution of the $m_\alpha$-resolved density matrix. Thus, we
have to calculate $\langle m_1, m_2 \lvert \partial_t
\rho _{\mathrm{osc}} \rvert m_1, m_2 \rangle$. We use
the relations
\begin{eqnarray*}
\langle m_1 , m_2 \lvert Y_1 Y^\dagger_1 \rho_{\mathrm{osc}}(t) \rvert
m_1 , m_2 \rangle &=& \rho (m_1 , m_2 ;t) \;,\\ \langle m_1 , m_2
\lvert Y^\dagger_1 \rho_{\mathrm{osc}}(t) Y_1 \rvert m_1 , m_2 \rangle
&=& \rho (m_1 -1, m_2 ;t) \;,\\ \langle m_1 ,m_2 \lvert Y^\dagger_1 Y_1
\rho_{\mathrm{osc}}(t) \rvert m_1, m_2 \rangle &=& \rho (m_1 , m_2 ;t)
\;,\\ \langle m_1, m_2 \lvert Y_1 \rho_{\mathrm{osc}}(t)
Y^\dagger_1\rvert m_1, m_2 \rangle &=& \rho (m_1 +1, m_2 ;t)\;,
\end{eqnarray*}
as well as the equivalent identities for detector 2 in Eq.~(\ref{eq
motion 2 detector setup}) to find the equation of motion for
$\rho(m_1,m_2;t)$. 

Following a counting-statistics approach\cite{belzig2003,blanter2006}, it is
particularly useful to express the equation of motion in terms of a
counting field $\chi_\alpha$, the conjugate quantity to the transfered
charge $m_\alpha$. Indeed, Fourier-transforming in the
transfered-charge indices $m_\alpha$,
\begin{align}
\tilde{\rho} (\chi_1,\chi_2;t) = \sum _{m_1=-\infty}^\infty e^{i\chi_1
m_1} \sum _{m_2=-\infty}^\infty e^{i \chi_2 m_2} \rho
(m_1,m_2;t)\end{align} leads to an equation of motion from which the
time-dependence of all moments of $m$ (for example, $\partial _t
\langle m_\alpha \rangle, \partial _t \langle m_\alpha^2 \rangle,
...$) can be determined. 
The current-current correlations can then be obtained
by taking successive derivatives with
respect to $(i \chi_\alpha)$ of the equation of motion of
$\tilde{\rho}(\chi_1,\chi_2;t)$.

In the regime of weak coupling between the oscillator and the
detectors, 
we can write the equation of motion of $\tilde{\rho}
(\chi_1,\chi_2;t)$ as
\begin{align}
\begin{split}\label{equation of motion rho}
&\frac{d}{dt} \tilde{\rho}(\chi_1,\chi_2;t) = \frac{-i}{\hbar} \left [
H_{\mathrm{osc}}, \tilde{\rho}(t) \right] + \frac{i}{\hbar} \sum
_\alpha\left[ \bar{F}_\alpha \hat{x} , \tilde{\rho}(t) \right] \\ &-
\frac{i}{\hbar} \sum_{\sigma,\alpha} \gamma_{\sigma,\alpha}
[\hat{x}, \{ \hat{p}, \tilde{\rho}(t) \}] - \frac{1}{\hbar^2}
\sum_{\sigma,\alpha} D_{\sigma,\alpha}
[\hat{x},[\hat{x},\tilde{\rho}(t)]] \\ &+\sum_{\sigma,\alpha} \left(
\frac{e^{i \sigma \chi_\alpha}-1}{\tau_{1,\alpha}^2} \right) \times \\
&\Bigg( \frac{2 D_{\sigma,\alpha}}{\hbar^2} ( \tau_{0,\alpha} +
e^{i \sigma \eta_\alpha} \tau_{1,\alpha} \hat{x} )\tilde{\rho}(t)(
\tau_{0,\alpha} + e^{-i \sigma \eta_\alpha} \tau_{1,\alpha} \hat{x} )
\\ &+i \frac{\gamma_{\sigma,\alpha}}{\hbar} \Bigl[ \tau_{0,\alpha}
\tau_{1,\alpha} (e^{i \sigma \eta_\alpha} \hat{p}\tilde{\rho}(t) -
e^{- i \sigma \eta_\alpha} \tilde{\rho}(t)\hat{p} ) \\ &+
 \tau_{1,\alpha} ^2 (\hat{p}\tilde{\rho}(t) \hat{x} -
\hat{x}\tilde{\rho}(t)\hat{p} ) \Bigr] \Bigg) \;.
\end{split}
\end{align}
Since $\tilde{\rho} (\chi_1=0,\chi_2=0;t) = \sum _{m_1} \sum _{m_2}
\rho(m_1,m_2;t)$, taking $\chi_1 = \chi_2 =0$ corresponds to
completely tracing out the charge degrees of freedom. In this case,
one finds that $\dot{\tilde{\rho}} (0,0;t)$ is of Caldeira-Leggett
form\cite{caldeira1983,caldeira1983a}. We can thus identify the constants
$D_{\sigma,\alpha}$ and $\gamma_{\sigma,\alpha}$ as, respectively, the diffusion and
damping constants induced by forward ($\sigma = +$) or backward ($\sigma = -$)
propagating currents in detector $\alpha$. We can also identify
$\bar{F}_\alpha$ as the average back-action force exerted on the
oscillator by detector $\alpha$. We find explicitly
\begin{align}
D_{\sigma,\alpha} &= \frac{\hbar^2}{4} \left(
\frac{\tau_{1,\alpha}}{\tau_{0,\alpha}} \right) ^2 \left[
\Gamma_{\sigma,\alpha} (\hbar \Omega) + \Gamma_{\sigma,\alpha}
(-\hbar \Omega) \right]\;,\\ \gamma_{\sigma,\alpha} &=
\frac{\hbar}{2M \Omega} \left( \frac{\tau_{1,\alpha}}{\tau_{0,\alpha}}
\right) ^2 \left( \frac{\Gamma_{\sigma,\alpha} (\hbar \Omega) -
\Gamma_{\sigma,\alpha} (-\hbar \Omega)}{2} \right) \;, \\
\bar{F}_\alpha&= \frac{1}{\hbar} \sin(\eta_\alpha) \left(
\frac{\tau_{0,\alpha}}{\tau_{1,\alpha}} \right) \sum_\sigma 2 \sigma
D_{\sigma,\alpha} \; ,
\end{align}
where the two inelastic tunneling rates are given by
\begin{eqnarray}
\Gamma_{+,\alpha} (E) &=& \frac{ \lvert \tau_{0,\alpha} \rvert^2}{h}
\label{gamma+} \\ 
&\times& \int _0 ^\infty d\varepsilon_{q,\alpha} \, f_{q,\alpha}
(\varepsilon_{q,\alpha}) \Bigl( 1- f_{k,\alpha}
(\varepsilon_{q,\alpha}+E) \Bigr) ,\nonumber \\ 
\Gamma_{-,\alpha} (E) &=&
\frac{\lvert \tau_{0,\alpha} \rvert^2}{h} \label{gamma-} \\
&\times& \int _0 ^\infty
d\varepsilon_{k,\alpha} \, f_{k,\alpha} (\varepsilon_{k,\alpha}) \Bigl(
1- f_{q,\alpha} (\varepsilon_{k,\alpha} + E) \Bigr) ,\nonumber
\end{eqnarray}
involving a transfer of energy $E$ from the oscillator to the lead
electron. We denote by $\Gamma_{+,\alpha}$ the forward 
tunneling rate, i.e. the rate at which electrons tunnel in the direction
favored by the voltage bias. The backward rate $\Gamma_{-,\alpha}$ corresponds to
the reverse process. In Eqs.~(\ref{gamma+}) and (\ref{gamma-}),
$f_{k,\alpha}=f_{R,\alpha}(\epsilon_{k,\alpha})$ is the Fermi distribution
function describing the local thermal equilibrium of the right lead of
detector $\alpha$ and $f_{q,\alpha}=f_{L,\alpha}(\epsilon_{q,\alpha})$ is the
same for the left lead.

Comparing these relations with the one derived in the
single-detector case\cite{clerk2004b} shows that the full damping and diffusion
coefficients governing the evolution of the oscillator are the sum of
two single-detector contributions.

The Caldeira-Leggett form of Eq.~(\ref{equation of motion rho}) allows
us to include the effect of direct coupling of the oscillator to the
environment by adding detector-independent contributions $D_0=2 M
\gamma_0 k_B T_{\mathrm{0}}$ and $\gamma_0 = \Omega / Q_0 $ (where
$Q_0$ is the extrinsic quality factor of the mode) to the previously
derived diffusion and damping constants. The evolution of the
oscillator is then governed by the two constants $D_{\mathrm{tot}}=
D_0 + \sum _{\sigma,\alpha} D_{\sigma,\alpha}$ and
$\gamma_{\mathrm{tot}}=\gamma_0 + \sum _{\sigma,\alpha}
\gamma_{\sigma,\alpha}$. For the specific case where the electronic
temperature is zero and where $eV_\alpha \gg \hbar \Omega$, current will only be
possible along the ($\sigma = +$) direction, and both $\gamma_{-,\alpha}$ and
$D_{-,\alpha}$ will be zero. In this case one can also show that
$\gamma_{+,\alpha} = \hbar \tau_{1,\alpha}^2 / (4\pi M)$ and that the
diffusion parameters are given by $D_{+,\alpha}=M \gamma_{+,\alpha}
eV_{\alpha}$. 

The equation of motion for different moments $\langle x^j p^k \rangle$
of the oscillator can be evaluated by taking the trace of $x^j p^k
\dot{\tilde{\rho}} (0,0;t)$. More generally, equations of motion for
combined moments of charge and oscillator quantities can be obtained
by also considering derivatives with respect to the counting fields
$\chi_\alpha$
\begin{align} \label{eq motion moments}
\begin{split}
\frac{\partial}{\partial t} &\langle x^{n_1} p^{n_2} m_1 ^{n_3}
m_2^{n_4} \rangle \\&= \mathrm{Tr} \; x^{n_1} p^{n_2} \left(
\frac{\partial^{(n_3+n_4)}}{\partial (i \chi_1)^{n_3} \partial (i
\chi_2)^{n_4}} \dot{\tilde{\rho}} (\chi_1,\chi_2;t) \right )
_{\chi_1=\chi_2=0} \;.
\end{split}\end{align}

\section{Single-detector case: Bound on the peak-to-background ratio}
\label{sec3}

One of the main motivations for studying position measurements using
cross-correlated detector outputs is to remove the bound on the
peak-to-background ratio that appears in the single-detector case,
just like in the case of a weak measurement of a two-level
system\cite{jordan2005}. In 
this section, we first review the results of Clerk and Girvin (CG)
[\onlinecite{clerk2004b}] for the single-detector configuration, in
the case where one considers the dc-biased, $T=0$, tunnel junction
where the $x-$dependent tunneling phase is $\eta=0$. We then carefully
analyze the peak-to-background ratio and show that this quantity is
bounded from above in the single-detector case, for finite bias voltage and oscillator displacement.

Using the single-detector analogue of Eq.~(\ref{equation of motion
rho}), CG showed that, under the conditions mentioned above and to
first non-vanishing order in $\tau_1$, the current noise of a tunnel
junction position detector is given by
\begin{align} \label{full noise one detector}
S_{I}^{\mathrm{tot}}(\omega) = 2 e \langle I \rangle + \frac{e^3 V}{h} (2 \tau_0
\tau_1)^2 \left( \frac{eV}{h} - \frac{\Omega}{4 \pi} \frac{\Delta
x_0^2}{\langle x^2 \rangle} \right) S_x (\omega) \;,
\end{align} 
where $\Delta x_0^2 = \hbar / (2 M \Omega)$ is the average of $x^2$ in the
ground state of the (quantum) harmonic oscillator and 
\begin{align}S_x (\omega) = 
\frac{8 \gamma_{\mathrm{tot}} \Omega^2 \langle x^2 \rangle}{ 4
\gamma_{\mathrm{tot}}^2 \omega^2 + (\Omega^2 - \omega^2)^2}\end{align} its power
spectrum. The full current noise is the sum
of the usual frequency-independent Poissonian (shot) noise and the
contribution of interest due to the coupling of the junction to
the oscillator. This second part is itself expressed as the difference
of a classical part (which is proportional to $V^2$) and a quantum correction
(which is proportional to $V$).

A relevant figure of merit of such detectors is the
peak-to-background ratio $\mathcal{R}(\omega)$: the ratio of the
contribution of the oscillator to the full current noise at frequency
$\omega$ over the unavoidable frequency-independent intrinsic detector
noise. This ratio is maximal at $\omega=\Omega$ and, in the case where
one only considers the $\propto V^2$ contribution in Eq.~(\ref{full
noise one detector}), was shown to be given by
\begin{align}
\mathcal{R}(\Omega) = \frac{S_{I}^{tot}(\Omega) - 2 e \langle I \rangle
}{2 e \langle I \rangle} = 4\tau_0^2 \frac{eV}{h \gamma_{tot}}
\frac{\beta^2}{1+\beta^2} \;,
\end{align}
where we used $\langle I \rangle = \langle \partial_t m_1(t) \rangle
\simeq e^2V \tau_0^2 \left( 1 + \beta^2 \right) / h $ and introduced
the dimensionless sensitivity parameter $\beta^2 = \tau_1^2 \langle
x^2 \rangle / \tau_0^2$.\cite{foot_sens} At this point, one should proceed with care
when maximizing $\mathcal{R}$ with respect to the sensitivity parameter, as
$\gamma_{\mathrm{tot}} = \gamma_0 + \gamma_+$ depends on $\beta$ through
$\gamma_+ = (\Omega \tau_0^2 / 2\pi ) (\Delta x_0^2/ \langle x^2
\rangle) \beta^2$. Writing out explicitely all terms in $\mathcal{R}$
that depend on $\beta$, one finds that
\begin{align}
\mathcal{R}(\Omega) = \frac{2 \tau_0^2}{\pi} Q_0 \frac{eV}{\hbar
\Omega} \left( 1+ Q_0 \frac{\tau_0^2}{2\pi}  \frac{\Delta
x_0^2}{\langle x^2 \rangle}  \beta^2 \right) ^{-1}
\frac{\beta^2}{1+\beta^2}
\end{align}
is a non-monotonic function of the sensitivity parameter $\beta$. For a given $\langle x^2 \rangle$, one can then
find an optimal value
\begin{align}\label{opt sens}
\beta^4_{opt} = \frac{2\pi}{Q_0 \tau_0^2} \frac{\langle x^2
\rangle}{\Delta x_0^2} \;,
\end{align}
for which $\mathcal{R}$ is maximal
\begin{align}
\mathcal{R}_{max} = 4 \frac{ Q_0 \tau_0^2}{2 \pi} \left(
\frac{eV}{\hbar \Omega}\right) \left(1+ \sqrt{\frac{Q_0
\tau_0^2}{2\pi} \frac{\Delta x_0^2}{\langle x^2 \rangle}
} \right) ^{-2} \;.
\end{align}

We can examine this result in two different limits. The first is when
the damping is mainly detector-independent ($\gamma_0 \gg \gamma_+$),
like in the case where the extrinsic quality factor of the resonator
is low, $Q_0 \ll \langle x^2 \rangle / ( \tau_0^2 \Delta x_0^2)$. In
this case, the maximal peak-to-background ratio,
\begin{eqnarray} \label{rmax low q}
\mathcal{R} &\simeq4& \frac{\langle x^2 \rangle}{\Delta x_0^2}
\frac{eV}{\hbar \Omega} \left( \frac{ \tau_0^2 Q_0}{2\pi} \frac{\Delta
x_0^2}{\langle x^2 \rangle}\right) \left( \frac{\beta^2}{1+\beta^2}
\right) \nonumber \\ &\le& 4 \frac{\langle x^2 \rangle}{\Delta
x_0^2} \frac{eV}{\hbar \Omega} \left( \frac{ \tau_0^2 Q_0}{2\pi}
\frac{\Delta x_0^2}{\langle x^2 \rangle}\right) \;,
\end{eqnarray} 
is reached when the sensitivity parameter $\beta$ is extremely large. However,
since the rightmost term of Eq.~(\ref{rmax low q}) is by definition 
small in this limit, the peak-to-background ratio
cannot become extremely large when the extrinsic resonator damping
dominates the detector-induced one.

Indeed, the real maximum of $\mathcal{R}$ is reached when one
considers the opposite limit of a very high resonator
$Q$-factor\cite{clerk2004b}, $Q_0 \gg \langle x^2 \rangle / ( \tau_0^2
\Delta x_0^2)$. For $\gamma_0 = 0$, the peak-to-background ratio can
be shown to obey
\begin{align}\label{max R}
\mathcal{R} \simeq 4 \frac{\langle x^2 \rangle} {\Delta x_0^2}
\frac{eV}{\hbar \Omega} \frac{1}{1+\beta^2} \le 4 \frac{\langle x^2
\rangle} {\Delta x_0^2} \frac{eV}{\hbar \Omega} \;.
\end{align}
In the single-detector case and for given system parameters ($eV$ and $\langle
x^2\rangle$), the peak-to-background ratio is therefore 
always bounded whatever the strength of the coupling and the bound
does not depend on $Q_0$ and $\tau_0$. As can be seen from Eq.~(\ref{opt
  sens}), the peak-to-background ratio is in this second 
case maximal in the limit $\beta \to 0$ of vanishing coupling. While the
optimal $\mathcal{R}$ can be increased by increasing the bias voltage, we
stress that our bound on $\mathcal{R}$ denotes the optimal value of the
peak-to-background reachable for a set of fixed system parameters.  

The nature of the true bound on $\mathcal{R}$ (i.e., the one found in
the case $Q_0 \to \infty$) is very similar to the Korotkov-Averin
bound that arises in the context of a weak measurements of a qubit. To
make this more apparent, we can derive this bound following the
linear-response approach that has been used to derive the bound on
$\mathcal{R}$ in the measurement of two-levels systems, treating the
detector as a position-to-current linear amplifier with responsivity (dimensionful gain) $\lambda
= 2 e^2 V \tau_0 \tau_1 / h$. As noted by CG, considering only the
dominant $\propto V^2$ term in Eq.~(\ref{full noise one detector})
corresponds to writing $\Delta S_I = S_I - 2 e \langle I \rangle=
\lambda^2 S_x (\omega)$. At resonance, the power spectrum $\Delta S_I
= 2 \lambda^2 \langle x^2 \rangle / \gamma$ is inversely proportional
to the damping rate $\gamma$, in the same way that the response of
the detector measuring a qubit is inversely proportional to the
dephasing rate due to the measurement device. Moreover, in both cases
one can show that the dephasing (damping rate) is proportional to the
fluctuations of the bare input of the detectors. For a position
detector in the high effective temperature limit $k_B T_{\mathrm{eff}}
\gg \hbar \Omega$, the detector-induced damping is indeed proportional
to the symmetrized detector force noise\cite{clerk2004a} $\gamma =
\overline{S}_F/ 2 M k_B T_{\mathrm{eff}}$, such that $\Delta
S_I \le 4 M \lambda^2 \langle x^2 \rangle k_B T_{\mathrm{eff}} /
\overline{S}_F$. Also, since for a tunnel junction detector there is
no reverse gain $\lambda^\prime$ and the real part of the
cross-correlator $\overline{S}_{IF}(\omega)$ vanishes\cite{clerk2004b}, the
condition on quantum-limited efficiency of the position
measurement\cite{clerk2004a} \begin{align} \label{quantum limit}
\overline{S}_I \overline{S}_F \ge \frac{\hbar^2}{4} (\mathrm{Re}[
\lambda - \lambda^\prime])^2 + (\mathrm{Re} [\overline{S}_{IF}])^2
\end{align}
becomes exactly the one used to derive the Korotkov-Averin bound
$\overline{S}_I \overline{S}_F \ge \frac{\hbar^2 \lambda^2}{4}$. We
then find that $ \mathcal{R} = \Delta S_I / S_I \le 8 \langle x^2
\rangle k_B T_{\mathrm{eff}} / (\hbar \Omega \Delta x_0^2)$. Using
$k_B T_{\mathrm{eff}} = eV/ 2 $ in the tunnel junction system, this
result corresponds exactly to Eq.~(\ref{max R}), the bound previously
derived using the equation-of-motion approach.

\section{Peak-to-background ratio in current cross correlations}
\label{sec4}

Extending ideas from the qubit measurement problem\cite{jordan2005}, we now
demonstrate 
how to eliminate the bound on the peak-to-background ratio in a
position measurement. Calculating the current-current correlations
between two tunnel-junction position detectors, we show that for
cross correlation measurements, $\mathcal{R}$ diverges. We also
obtain analytical results for the cross correlations in two typical
cases.

To calculate the current cross correlations, we use the generalized
MacDonald formula\cite{rodrigues2005,macdonald1962}, a general result (valid for stationary processes) that provides a way, in the present case, to relate the symmetrized cross correlations to the Fourier sine-transform of the time-derivative of the covariance of $m_1$ and $m_2$, the number of charges that tunneled through each junction. The generalized MacDonald formula reads
\begin{align}
\label{MacDonald cross correlations}
S_{I_1,I_2} (\omega) = 2 e^2 \omega &\int _0 ^\infty dt \sin(\omega
t) K_{1,2}(t) \;,
\end{align}
where we defined
\begin{align}
K_{1,2} (t) = \Bigr[   \frac{d}{dt^\prime} \big( \langle m_1  m_2  \rangle_{t^\prime} - \langle m_1 \rangle_{t^\prime} \langle m_2  \rangle_{t^\prime} \big) \Bigr]_{t^\prime=t} \;.
\end{align}
In this last equation $\langle m_1 m_2 \rangle_t$ corresponds to $\mathrm{Tr} \,m_1\, m_2 \,\tilde{\rho}(0,0,t)$ and represents the coupled moment of $m_1$ and $m_2$ at time $t$. 

To proceed further, we restrict
ourselves to the case of zero electronic temperature and dc-bias. In the
following subsections, we analyze in detail the two different cases
depicted schematically in Fig.~\ref{Setup}. We have in mind that a realization
of the setup shown in Fig.~\ref{Setup} is made in a similar way as the
single-detector setup in
Ref.~\onlinecite{flowers-jacobs2007}. This means that the tunnel junctions
correspond to atomic point contacts (formed by electromigration) which are
separated by about 1 nm from the oscillator. In contrast, the two detectors are
assumed to be separated from each other by at least 20 nm. Therefore,
capacitive cross-talking between the detectors will play a negligible role. \cite{foot_cap}

\subsection{In-phase configuration}\label{case 1}

We will first consider the case where both $\eta_1 = \eta_2 =0$,  the case
where both tunnel 
junctions are located on the same side of the oscillator, {\it cf.}
Fig.~\ref{Setup} a). To calculate the cross correlations, we use Eq.~(\ref{eq motion moments}) (with $n_1=n_2=0$), to find that 
\begin{align*}
&\frac{d}{dt} \langle m_\alpha  \rangle_t = \frac{eV_ \alpha}{h} \left( \tau_{0, \alpha}^2 + 2  \tau_{0, \alpha} \tau_{1, \alpha}  \langle x \rangle + \tau_{1_ \alpha}^2 \langle x ^2 \rangle   \right) - \gamma_{+,\alpha}\;,\\
\begin{split}
\frac{d}{dt}& \langle m_1 m_2 \rangle_t =\\ &\frac{eV_1}{h} \left( \tau_{0,1}^2 \langle m_2 \rangle_t  + 2  \tau_{0,1} \tau_{1,1} \langle x m_2 \rangle_t  + \tau_{1,1}^2 \langle x ^2 m_2\rangle_t    \right)  \\
+ &\frac{eV_2}{h} \left( \tau_{0,2}^2 \langle m_1 \rangle_t + 2  \tau_{0,2} \tau_{1,2}\langle x m_1\rangle_t  + \tau_{1,2}^2 \langle x ^2 m_1\rangle_t    \right) \\  - &\gamma_{+,1} \langle m_2 \rangle_t  -  \gamma_{+,2} \langle m_1 \rangle_t \;,  \end{split}
\end{align*}
and therefore that $K_{1,2}(t)$ in this case is given by
\begin{align} \begin{split}
K_{1,2}(t) = &2 \frac{eV_1}{h} \tau_{0,1}\tau_{1,1} \langle \langle xm_2\rangle \rangle_t  +  \frac{eV_1}{h} \tau_{1,1}^2 \langle \langle x^2 m_2 \rangle \rangle_t  \\ + &2\frac{eV_2}{h} \tau_{0,2}\tau_{1,2} \langle \langle xm_1\rangle \rangle_t +  \frac{eV_2}{h} \tau_{1,2}^2 \langle \langle x^2 m_1 \rangle \rangle_t\;.
\end{split}
\end{align}
where the double bracket denotes the covariance of two quantities:
$\langle \langle a b \rangle \rangle_t \equiv \langle a b \rangle_t -
\langle a \rangle_t \langle b \rangle_t $. This means that, to lowest order in $\tau_{1,\alpha}$, the \emph{full} cross-correlated output of the
detectors is given in this
configuration by
\begin{align}\label{cross-co xm1 xm2}
\begin{split}
&S_{I_1,I_2} (\omega)\Big \lvert_{\substack{\eta_1 = 0 \\ \eta_2 = 0}}
= 4 e^2 \omega \int _0 ^\infty \,dt \sin (\omega t) \times \\ &\left(
\frac{e V_1}{h} \tau_{0,1} \tau_{1,1} \langle \langle x m_2 \rangle
\rangle_t + \frac{e V_2}{h} \tau_{0,2} \tau_{1,2} \langle \langle x
m_1 \rangle \rangle_t \right)\;.
\end{split}
\end{align}

The cross-correlated signal does not contain any oscillator-independent contribution. Using Eq.~(\ref{eq motion moments}), a closed system of differential equations involving $\langle \langle pm_\alpha \rangle \rangle _t$ and $\langle \langle xm_\alpha \rangle \rangle _t$ can be
generated. This system can be solved, using the boundary conditions
$m_\alpha(0)=0$ and assuming that all averages that do not contain
$m_\alpha$ are time-independent and can therefore be evaluated in the stationary
($t\to \infty$) limit\cite{armour2004}.

Solving for the different covariances, we find that the current
cross correlations can be written as
\begin{align}\label{cross correlations deltaeta zero}
\begin{split}
S_{I_1 I_2}^{\mathrm{tot}}(\omega) =& e^2 (2 \tau_{0,1}\tau_{0,2}) (2
\tau_{1,1}\tau_{1,2}) \\ &\left( \frac{e^2 V_{1} V_{2}}{h^2} -
\frac{e(V_1+V_2) }{2 h} \frac{\Omega}{4 \pi} \frac{\Delta
x_0^2}{\langle x^2 \rangle} \right) S_x (\omega) \;,
\end{split}\\
&= \lambda_1 \lambda_2 \left( 1-  \frac{\hbar \Omega (eV_1 + eV_2)}{4 eV_1 eV_2}\frac{\Delta x_0^2}{\langle x^2 \rangle} \right) S_x (\omega) \;,
\end{align}
where we introduced the gains $\lambda_\alpha = 2 e^2 \tau_{0,\alpha} \tau_{1,\alpha} V_\alpha \cos (\eta_\alpha) / h$.
Evidently, the cross-correlated output of
the detectors (\ref{cross correlations deltaeta zero}) does not contain 
any frequency-independent
background noise. The peak-to-background ratio $\mathcal{R} (\Omega)$
therefore diverges for all values of $\gamma_0 / \gamma_+$, not
because of an increased signal but due to the absence of background
noise in this configuration.

For this type of measurement, a relevant
figure of merit of the detection system $\mathcal{R}_c$ is the ratio
of the cross-correlated output over the frequency-independent noise
power of individual detectors : $\mathcal{R}_c = S_{I_1,I_2}^{\mathrm{tot}}(\Omega)/
\sqrt{S_1 S_2}$, where $S_\alpha = 2 e \langle I_\alpha \rangle$. For
our position detector, we find
\begin{widetext} 
\begin{align}
\mathcal{R}_c = \frac{\lvert S_{I_1,I_2} ^{\mathrm{tot}}\rvert }{\sqrt{S_1 S_2}} &=
\frac{4}{ 1 + \frac{\gamma_0}{ \gamma_+}}
\frac{1}{\sqrt{(1+\beta_1^2)(1+\beta_2^2)}} \frac{ \tau_{1_1}
\tau_{1,2} }{\tau_{1,1} ^2 + \tau_{1,2} ^2} \frac{\sqrt{V_1
V_2}}{V_1+V_2} \frac{e (V_1+V_2)}{\hbar \Omega}\frac{\langle x^2
\rangle}{\Delta x_0^2} \le \frac{e (V_1+V_2)}{\hbar
\Omega}\frac{\langle x^2 \rangle}{\Delta x_0^2} \;,
\end{align}
\end{widetext}
where we used $2 xy \le ( x^2 + y^2)$. From this inequality, we see
that the maximal cross-correlated output is found for (i)
twin-detectors (where $\tau_{1,1} = \tau_{1,2}$) and (ii) equal bias
voltages $V_1=V_2$. Also, like in the single-detector case,
$\mathcal{R}_{c}$ is maximal in the limit where there is no extrinsic
oscillator damping $\gamma_0$ and where the correction to the average
current due to the coupling to the oscillator vanishes ($\beta_\alpha
\to 0$).

Once again it is instructive to compare our value of $\mathcal{R}_c$
for twin detectors with the equivalent result in the case of a weak
measurement of a qubit using cross correlations\cite{jordan2005}. In
the latter case, the cross-correlated output was shown to be limited
to $1/2$ of the single-detector signal due the increased (doubled)
detector-induced dephasing. This is the same here.

\subsection{Out-of-phase detection} \label{case 2}

We can also analyze the case where one detector couples to $+x$ and
the other to $-x$, as would happen if the two detectors were located
on opposite sides of the resonator (see Fig.~\ref{Setup}). In terms of the
tunneling phases $\eta_\alpha$, this corresponds to taking $\eta_1=0$
and $\eta_2 = \pi$. Using Eq.~(\ref{MacDonald cross correlations}),
the cross correlations are then given by
\begin{align}\label{cross-co xm1 xm2 deltaetapi}
\begin{split}
&S_{I_1,I_2} (\omega)\Big \lvert_{\substack{\eta_1 = 0 \\ \eta_2 =
\pi}} = 4 e^2 \omega \int _0 ^\infty \,dt \sin (\omega t) \times \\
&\left( \frac{e V_1}{h} \tau_{0,1} \tau_{1,1} \langle \langle x m_2
\rangle \rangle_t - \frac{e V_2}{h} \tau_{0,2} \tau_{1,2} \langle
\langle x m_1 \rangle \rangle_t \right) .
\end{split}
\end{align}
As the coupling between detector 1 and the oscillator is the same as
in the previous case $\langle \langle x m_1 \rangle \rangle_t$ remains
unchanged in this second configuration. The covariance $\langle
\langle x m_2 \rangle \rangle_t$ on the other hand changes sign (but
keeps the same norm) in this new configuration. Equation
(\ref{cross-co xm1 xm2 deltaetapi}) then yields
\begin{align}\label{cross-co type 2} S_{I_1,I_2} &(\omega)\Big
\lvert_{\substack{\eta_1 = 0 \\ \eta_2 = \pi}} = -
S_{I_1,I_2}(\omega)\Big \lvert_{\substack{\eta_1 = 0 \\ \eta_2 = 0}}
\;.
\end{align}
The cross correlations in the second configuration are the same as in
the first one, but of negative sign. From an amplifier point of view,
this is easily explained since putting $\eta_2 = \pi$ corresponds to
transforming $\lambda_2 \to - \lambda_2$ in $S_{I_1,I_2} \simeq \lambda_1
\lambda_2 S_x$. Finally, note that this configuration was analyzed for
two single-electron transistor position detectors coupled to a
classical oscillator, in Ref. [\onlinecite{rodrigues2005}] by
Rodrigues and Armour. In their article, these
authors only explicitly calculated zero-frequency cross correlations
between the currents in both detectors, but they conjectured that, at the
resonance frequency of the oscillator, this detector-configuration
(corresponding to $\eta_1=0,\eta_2=\pi$ in our approach) should yield
strong negative cross correlations, just like the ones predicted
here. 

\section{Bound on the added displacement noise}
\label{sec5}

As shown in Sec. \ref{sec3}, to derive the equivalent of the Korotkov-Averin
bound in a position measurement, one needs to consider the \emph{full} current
noise, where no distinction is made between the signal due to the intrinsic
equilibrium fluctuations of the oscillator $S_I^{\mathrm{eq}} (\omega)$ and the
remainder of the signal $S_{I}^{\mathrm{add}} (\omega)$. This second contribution contains, amongst other things, the added signal due to heating
of the oscillator by the detector. When trying to measure precisely the equilibrium fluctuations of a nanomechanical oscillator however, it is important to consider the two contributions separately: $S_I^{\mathrm{eq}} (\omega)$ is exactly what you would like to measure while $S_{I}^{\mathrm{add}} (\omega)$ limits the sensitivity of the measurement. When using a single linear detector like the tunnel junction, this measurement sensitivity is quantum-mechanically bounded from below\cite{clerk2004a}. 

When discussing this bound on added noise, one usually considers the added
\emph{displacement} noise, that corresponds to the added current noise
referred back to the oscillator. We therefore introduce the total displacement
noise $S_{x}^{\mathrm{tot}} $, defined as 
\begin{align}
S_{x}^{\mathrm{tot}} (\omega) = \frac{S_{I}^{\mathrm {tot}} (\omega) }{\lambda^2} =
S_{x}^{\mathrm{add}} (\omega) + S_{x}^{\mathrm{eq}} (\omega)\;, 
\end{align}
where $\lambda$ is the $x$-to-$I$ gain of the detector, $S_{x}^{\mathrm{add}} (\omega)$ is the part of the full
displacement spectrum that arises due to the presence of the detector. In the
relevant limit of a detector with a high power gain ($eV  \gg \hbar
\Omega$), it was shown using general arguments that $S_{x}^{\mathrm{add}}
(\Omega) \ge \hbar / M \Omega \gamma_{\mathrm{tot}}$: the best possible detector
therefore adds exactly as much noise as a zero-temperature bath of frequency
$\Omega$\cite{clerk2004a,caves1982}.  

Before discussing the limit on the added displacement noise in a cross correlation setup, it is helpful to describe how the quantum limit on $S_{x}^{\mathrm{add}} (\Omega)$ is reached in a single-detector configuration. Let's consider for definitiveness the experimentally relevant configuration where $eV \gg k_B T_0 > \hbar \Omega$. For a measurement to be quantum limited, the effective temperature of the oscillator $T_{\mathrm{eff}}=(\gamma_+ eV/2 + \gamma_0 k_B T_0)/(k_B \gamma_{\mathrm{tot}})$ must not be dramatically higher than $T_0$. This is natural, since added fluctuations due to the higher effective temperature are, by definition, unwanted back-action noise. In this regime, one therefore cannot expect $S_x^{\mathrm{add}}$ to be close to the quantum limit unless $\gamma_+ \ll \gamma_0$. The regime of $\gamma_+ / \gamma_0$ in which quantum-limited displacement sensitivity can be achieved is therefore very different from the one where the bound on the peak-to-background ratio can be reached.

Using the expression for the full current noise derived earlier (Eq. (\ref{full noise one detector})), we write the full position noise as
\begin{align}
S_{x}^{\mathrm{tot}}  (\omega)&=\frac{S_{I}^{\mathrm{tot}}(\omega)}{\lambda^2} = \frac{2 e \langle I \rangle}{\lambda^2} + \left(1-\frac{\hbar \Omega}{2 eV} \frac{\Delta x_0^2}{\langle x^2 \rangle} \right) S_x (\omega)\;,\\
=  \frac{2 e \langle I \rangle}{\lambda^2} &+ 8  M \gamma_{\mathrm{tot}} k_B T_{\mathrm{eff}}\lvert g(\omega)\rvert^ 2 - 2 M \gamma_{\mathrm{tot}} \frac{(\hbar \Omega)^2}{eV} \lvert g(\omega)\rvert^2\;,
\end{align}
where in the last line we introduced the oscillator's response function $g^{-1}(\omega) = M[(\Omega^2 - \omega^2) + 2 i \gamma_{\mathrm{tot}}  \omega]$. Splitting the second term into a detector dependent and independent part, we find
\begin{align}
S_{x}^{\mathrm{eq}} &= 8 M \gamma_0 k_B T_0 \lvert g(\omega) \rvert^2\;,\\
S_{x}^{\mathrm{add}}&= \frac{2 e \langle I \rangle}{\lambda^2} + 8  M \gamma_+ \frac{eV}{2}\lvert g(\omega)\rvert^2- 2 M \gamma_{\mathrm{tot}} \frac{(\hbar \Omega)^2}{eV} \lvert g(\omega)\rvert^2 \;.
\end{align}
This way of writing the equilibrium fluctuations implies that we consider $\gamma_{\mathrm{tot}} \simeq \gamma_0$ in $g(\omega)$, in agreement with our previous assumption that $\gamma_+ \ll \gamma_0$. The added noise contains three contributions, corresponding to the detector shot noise, the detector-induced heating of the oscillator and a correction ($\propto \hbar \Omega / eV$) arising from the cross correlation between the detector output noise and the back-action force, $\overline{S}_{IF} $, respectively. Explicitly, taking $\langle I \rangle \simeq  e^2 \tau_0^2 V / h $,\footnote{The derived bound is therefore valid up to a positive correction of order $\beta^2$.} we obtain
\begin{align}
S_{x}^{\mathrm{add}}&= \frac{\pi \hbar}{eV \tau_1^2} +\frac{ \hbar \tau_1^2 eV \lvert g(\omega)\rvert^2 }{\pi } - 2 M \gamma_{\mathrm{tot}} \frac{(\hbar \Omega)^2}{eV} \lvert g(\omega)\rvert^2 \;.
\end{align}
For a fixed bias voltage, the relevant tunable parameter is directly the detector-oscillator coupling $\tau_1$ (and not the dimensionless sensitivity parameter $\beta$, since $S_{x}^{\mathrm{add}}$ is independent of $\tau_0$).\footnote{In principle, we could use the bias voltage $eV$ as an optimization parameter. In this case, we would find that $S_{x}^{\mathrm{add}} \to 0$ for $eV/\hbar \Omega \to 0$; there is no limit on the added position noise in the low power gain regime ($eV \sim \hbar \Omega$) \cite{clerk2004b,caves1982}. However, since Eq.~(\ref{full noise one detector}) was derived in the high bias regime, it is better in the present case to optimize the coupling strength $\tau_1$ while keeping $eV/\hbar \Omega \gg 1$ fixed. } For strong coupling, $S_x^{\mathrm{add}}$  is dominated by heating of the oscillator, while for weak coupling, the shot noise contribution ($\propto 1/\tau_1^2$) dominates. This is the regime in which the current generation of experiments are operated \cite{flowers-jacobs2007}. There is an optimal coupling $ \tau_{1,\mathrm{opt}}^2 = \pi/( eV \lvert g(\omega) \rvert)$ 
that minimizes the total added noise. At the resonance, we recover the inequality
\begin{align}
S_{x}^{\mathrm{add}} (\Omega) \ge \left( 1 - \frac{\hbar \Omega}{2 eV} \right) \frac{\hbar}{\gamma_{\mathrm{tot}} M  \Omega}\;,
\end{align}
where the bound is reached when $\tau_1 = \tau_{1,\mathrm{opt}}$. This is the quantum limit on the added displacement noise for the single-detector configuration. In passing, we note that the effective temperature of the oscillator when the coupling strength $\tau_1$ is optimal is \begin{align}
T_{\mathrm{eff}} = T_0 + \frac{\hbar \Omega}{4 k_B} \;,
\end{align}
in agreement with the general analysis of Ref.~[\onlinecite{clerk2004a}].
The heating of the oscillator by the detector is, as expected, very low when doing a quantum-limited measurement. 

We can now show how cross correlations can be used to beat the quantum limit on $S_x^{\mathrm{add}}$ derived in the single-detector case. In both cross correlation configurations ($\eta=0,\pi$), $S_{x}^{\mathrm{tot}} = S_{I_1,I_2} / \lambda_1 \lambda_2$ is identical. Like in the single-detector case, we separate the total position fluctuations in two parts 
\begin{align}
S_{x}^{\mathrm{eq}} &= 8 M \gamma_0 k_B T_0 \lvert g(\omega) \rvert^2\;,\\
\frac{S_{x}^{\mathrm{add}}}{M \lvert g(\omega) \rvert^2}&=  4   \left( \sum _\alpha \gamma_{+,\alpha} e V_\alpha \right)-  \gamma_{\mathrm{tot}} \frac{(\hbar \Omega)^2(eV_1+eV_2)}{eV_1 eV_2}\;.
\end{align}
The cross-correlated position spectrum does not contain the frequency-independent shot noise contribution that diverges for low coupling ($\propto 1/\tau_1^2$). Therefore, one does not need to tune the coupling to equilibrate the ``shot noise'' and back-action ``heating'' contributions. Instead, one can freely tune the couplings $\tau_{1,\alpha}$ such that $S_{x}^{\mathrm{add}} (\omega)$ vanishes completely. We find $S_{x}^{\mathrm{add}}=0$ for $\tau^2_{1,\alpha,\mathrm{opt}} = 4 \pi M \gamma_{+,\alpha,\mathrm{opt}} / \hbar$, where
\begin{align}
\gamma_{+,\alpha,\mathrm{opt}}= \frac{\gamma_{\mathrm{tot}}}{4} \left( \frac{\hbar \Omega}{eV_\alpha} \right)^2\;.
\end{align}
At the optimal coupling point, the effective temperature of the oscillator is
\begin{align}
T_{\mathrm{eff}} = T_0 + \left( \frac{\hbar \Omega}{eV_1} + \frac{\hbar \Omega}{eV_2} \right) \frac{\hbar \Omega}{8 k_B}\;.
\end{align}
In the regime of interest ($eV_\alpha \gg \hbar \Omega$), the additional heating of the oscillator considerably reduced from the single-detector value.

\section{Conclusion}
In this article, we have shown that, for a tunnel-junction position
detector coupled to a nanomechanical oscillator, the optimal
peak-to-background ratio $\mathcal{R}$ at the resonance frequency of
the oscillator is bounded. In contrast to the universal (independent of all system parameters) bound derived for a
continuous weak measurement of qubits ($\mathcal{R} \le 4$), the new
bound derived for position measurements is a function of the effective
temperature of the oscillator and its average displacement. We have
also shown that adding a second detector and using the
cross correlations between the two detectors allows one to eliminate
this bound on $\mathcal{R}$. We have analyzed in detail the
cross-correlated output of the position detectors in two typical
configurations, and have shown that in both cases the optimal
cross-correlated signal is measured by twin detectors.  We also investigated the quantum-limit on the added displacement noise and shown that it is possible to totally eliminate the added displacement noise by doing a cross-correlated measurement. This configuration therefore opens the door for displacement measurement with sensitivities better than the quantum limit.

\section{Acknowledgments}
We would like to thank A.A. Clerk and A.N. Jordan for
interesting discussions and correspondence. This work was financially
supported by the Natural Sciences and Engineering Research Council of Canada,
the Fonds Qu\'eb\'ecois de la Recherche sur la Nature et les Technologies, the
Swiss NSF, and the NCCR Nanoscience.

\end{document}